\documentclass[superscriptaddress, amsmath, aps, prl, numerical, twocolumn, nofootinbib]{revtex4-1}

\usepackage[T1]{fontenc}
\usepackage{latexsym}
\usepackage{amssymb}
\usepackage{amsfonts}
\usepackage{amsmath}
\usepackage{amsthm}
\usepackage{multirow}
\usepackage{longtable}
\usepackage{mathtools}
\usepackage{color}
\usepackage{ulem}
\usepackage{hyperref}
\usepackage{bm}

\newcommand{\rf}{\langle r^4\rangle}
\newcommand{\rt}{\langle r^2\rangle}

\begin{document}

\title{Beyond the charge radius: the information content of the fourth radial moment}
    \author{P.-G.~Reinhard}
    \affiliation{Institut f\"{u}r Theoretische Physik, Universit\"{a}t Erlangen, D-91054 Erlangen, Germany}
          \author{W.~Nazarewicz}
    \affiliation{Department of Physics and Astronomy and FRIB Laboratory, Michigan State University, East Lansing, Michigan 48824, USA}
          \author{R.~F.~Garcia Ruiz}
    \affiliation{Physics Department, CERN, CH-1211 Geneva 23, Switzerland}
    \affiliation{Massachusetts Institute of Technology, Cambridge, MA 02139, USA}
  
\begin{abstract}
Measurements of atomic transitions in different isotopes offer key information on the nuclear charge radius. The anticipated high-precision experimental techniques, augmented by atomic calculations, will soon enable extraction of the  higher-order  radial moments of the  charge density distribution. To assess the value  of such measurements for nuclear structure research, we study the information content of the fourth radial moment $\rf$ by means of nuclear density functional theory and a multiple correlation analysis. We show that $\rf$ can be directly related to the surface thickness of nuclear density, a fundamental property of the atomic nucleus that is difficult to obtain for radioactive systems. Precise knowledge of these radial moments is essential to establish reliable constraints on the existence of new  forces from precision isotope shift measurements.
\end{abstract}
  
\maketitle

\textit{Introduction--}
A precise knowledge of the electron-nucleus interaction in atoms can provide access to physical phenomena relevant to a wide range of energy scales. High-precision measurements of atomic transitions, for example, offer complementary information to our understanding of the atomic nucleus, the study of fundamental symmetries, and the search for new physics beyond the Standard Model of particle physics \cite{Saf18,Ber18,Sta18,Del17}.

Varying the number of neutrons induces changes in the charge density distribution along the isotopic chain, causing tiny perturbations in the energies of their atomic electrons, known as isotope shifts. Measurements of the corresponding frequencies, typically of the order of MHz, allow changes in the root-mean-squared (rms) nuclear charge radii to be extracted \cite{Gar16,Cam16}. Extending these measurements for isotopes away from stability is of marked and growing interest for low-energy nuclear physics, as the data on the nuclear size are essential for our understanding of the nuclear many-body problem \cite{Gar16,Ham18,Marsh18,Gor19,Mil19}.
In recent years, the interest in precision isotope shift measurements has increased significantly. Performing measurements across long isotope chains that are readily available at state-of-the-art radioactive ion beam facilities has the potential to constrain the existence of new forces and hypothetical particles with unprecedented sensitivity \cite{Del17,Del17b,Fru17,Fla18,Ber18,Sta18}.
This has motivated the rapid progress of experimental techniques which are continuously pushing the frontiers of precision measurements. Quantum logic detection schemes have achieved sub-MHz precision \cite{Geb15}, and recent developments such as spin squeezing \cite{Bra19} and quantum entanglement \cite{Man19}, are now able to reach sub-Hz precision. This level of precision offers sensitivity not only to explore the new physics, but would also provide access to nuclear observables that have so far been elusive, such as  the higher-order radial moment  $\rf$ \cite{Pap16,Ekman2019} and the nuclear dipole polarizability  \cite{Fla18}.
Precise knowledge of these nuclear properties  will open up exciting opportunities in nuclear structure research; hence, is essential to  establish reliable constraints in the exploration of new physics \cite{Fla18}.
In addition to the progress of high-precision experiments, the continued development of atomic and nuclear theory has played a crucial role to extract nuclear structure and fundamental physics observables from measurements \cite{Pap16,Ekman2019}.

The isotope shift, $\Delta \nu_i $, between an isotope with mass, $A$, and an isotope ${A'}$, can be expressed by a product of nuclear and atomic factors as
\begin{eqnarray}
\Delta \nu_i^{AA'} &=& K_{\text{MS}, i}  \frac{A - {A'}}{A {A'}}+\sum_k F_{i,k} \delta \langle r^{2k} \rangle ,  
\label{IS}
\end{eqnarray}
where $\delta \langle r^{2k} \rangle $ is the difference between the nuclear radial moments of order $2k$. The atomic part is factorized in the constants $K_{\text{MS},i}$ and $F_i$, referred to  as the mass shift and the field shift, respectively. Assuming a negligible contribution from $k>1$ moments,  isotope shifts from different atomic transition  $i$ and $j$, $\Delta \nu^{AA'}_i$ vs $\Delta \nu^{AA'}_j$, should follow a linear relation known as the King plot \cite{king}. The non-linearity of the King plot can be due to  the contribution from $k>1$ moments. It can also indicate the presence of new phenomena  \cite{Del17,Del17b,Fru17,Fla18,Ber18,Sta18}. Therefore, the estimation of 
 the effect higher-order terms is important  to provide bounds on  physics beyond the Standard Model.
As discussed in Refs.~\cite{Pap16,Ekman2019}, by taking advantage of the improved experimental precision and  atomic calculations with well-controlled uncertainty quantification for
atomic states, it will enable us to extract highly accurate atomic line field shifts and higher-order radial moments.
To assess the impact of this new anticipated data on  our  understanding of atomic nuclei, in this Letter, we employ density functional theory to study  the $k=2$ moment  $\rf$ of nuclear charge distribution.

\textit{Nuclear charge distribution characteristics--}
The gross features of the nuclear charge distribution $\rho(\boldsymbol{r})$ and its charge form factor $F(\boldsymbol{q})$ can be described by form
parameters: radial moments 
\begin{equation}
  r_n \equiv\sqrt[n]{\langle r^n \rangle}
  =
  \left(
  \frac{\int d^3r\,r^n\,\rho(\boldsymbol{r})}{\int d^3r\,\rho(\boldsymbol{r})}
  \right)^{1/n},
\label{eq:rgeneral}
\end{equation}
diffraction radius $R$, and surface thickness $\sigma$. The latter characterizes the density diffuseness around the nuclear surface.
The rms charge radius is given by the second moment 
$r_2$.  The diffraction radius  is
determined from the first zero of the
form factor $F(q)$ (the diffraction minimum).
 It represents a box-equivalent
radius. The surface thickness is  determined from the height of
the first maximum of $F(q)$. The
relations between spatial geometrical parameters and the form factor are provided by
the Helm model \cite{Hel56aE}, which represents the nuclear density
profile by a folding of a box distributions (having radius $R$) with a
Gaussian of width $\sigma$. For details see \cite{Fri82a,Andrae2000} and the
supplemental material \cite{SM}. Within the  Helm model,  $r_2$ can be expressed in terms of
$R$, and $\sigma$: $r_2^{\rm (H)}=\sqrt{\frac{3}{5}R^2+3\sigma^2}$. In practice, this relation is not exactly fulfilled and 
the deviation characterizes the halo of the charge distribution \cite{Mizutori2000,Sarriguren07,Schunck08}:
\begin{equation}
  h  =  r_2-r_2^{\rm (H)}.
\label{eq:halo}
\end{equation}
In practice,
the halo is a small positive quantity \cite{Mizutori2000}. Diffused charge distributions associated with loosely bound protons in proton rich isotopes produce appreciable values of  $h$.
\begin{figure}
\centerline{\includegraphics[width=0.8\columnwidth]{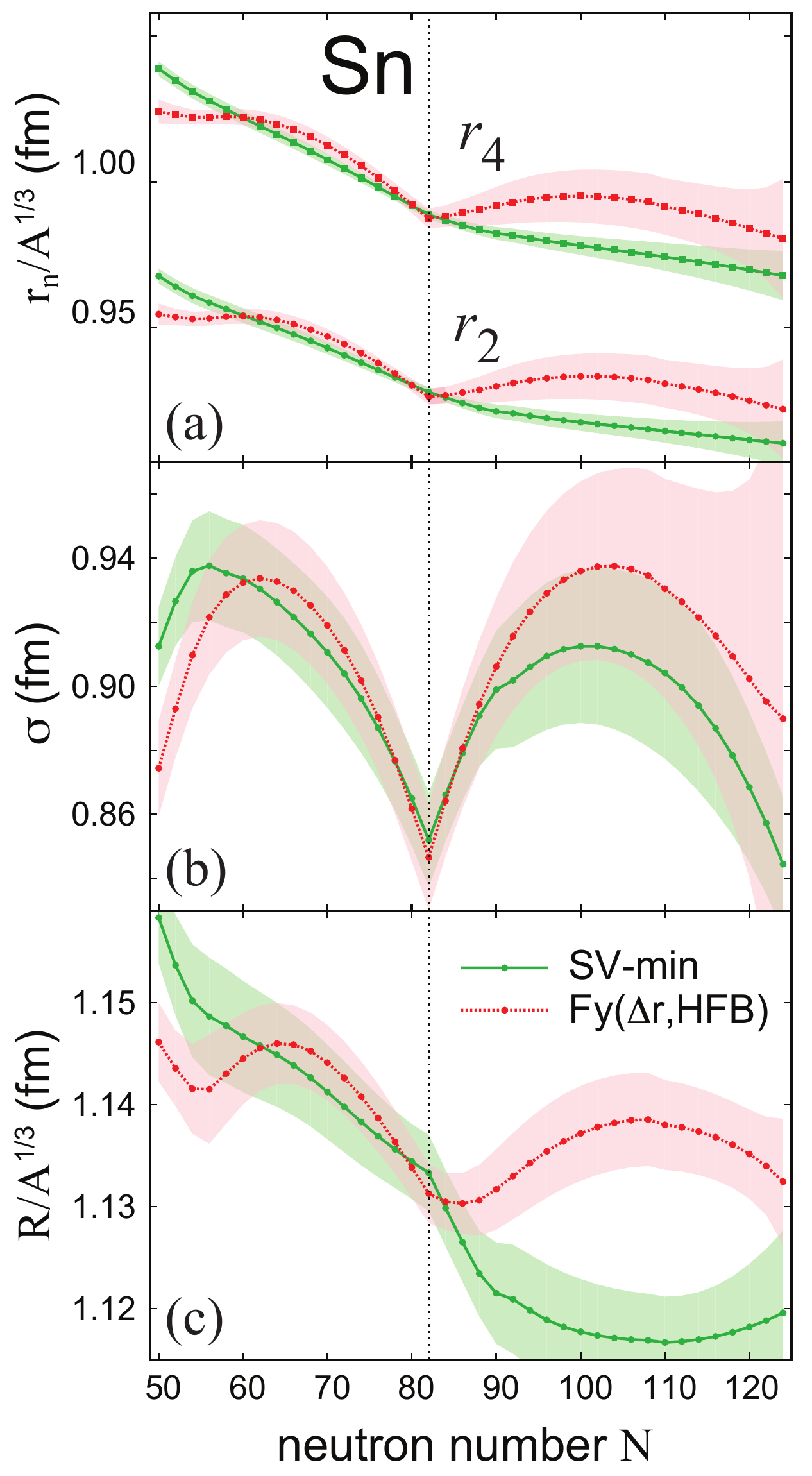}}
\caption{\label{fig:radii} Form parameters along the Sn chain ($Z=50$)
  computed with energy density functionals SV-min  and Fy($\Delta$r) 
  together with their statistical uncertainties.  Shown are 
radial moments $r_2$ and $r_4$ (a); surface thickness $\sigma$ (b); and
diffraction radii $R$  (c). To better visualize the local trends, radial moments and diffraction radii are scaled by $A^{1/3}$. The magic neutron number  $N=82$ is marked.
}
\end{figure}
To give an idea about the typical values and trends, we show in 
Fig.~\ref{fig:radii} the  charge density form parameters for the chain of Sn isotopes
calculated with  two nuclear energy density functionals: the Skyrme parametrization
SV-min \cite{Klu09a} and the Fayans functional Fy($\Delta r$,HFB)
\cite{Rei17a,Mil19,SM}. Both functionals have been optimized  with respect to the 
pool of empirical data from \cite{Klu09a}, with some additional charge radii data used 
in the optimization of Fy($\Delta r$,HFB). The fits allow also to
deduce the statistical uncertainties on the predicted observables by
standard linear regression methods \cite{Dob14a}. The uncertainties comply nicely with
the adopted errors for the observables which are $\pm 0.04$\,fm for $R$
and $\sigma$, and $\pm 0.02$\,fm for $r_2$. These errors do not contain
the experimental uncertainties but reflect the capability of the
model to reproduce observables. A measurement is expected to provide a   new information if the experimental uncertainty is safely below the model error.
The values of the form parameters show the expected trends
\cite{Mizutori2000}. Namely, the proton radii shrink systematically
with increasing neutron number because of the increasing proton
binding. The pronounced kink at the $N=82$ shell closure seen in all
form parameters is due to reduced neutron pairing \cite{Gor19}.  The
two density functionals used deliver similar results in the domain of
well bound nuclei while developing slight differences at exotic
proton-rich nuclei close to $N=50$ and neutron rich nuclei with
$N>82$. This is entirely anticipated: form parameters of stable
nuclei, being part of the optimization data pool, are bound to be well
reproduced.  The differences between SV-min and Fy($\Delta r$,HFB) in
neutron rich isotopes are almost exclusively generated by the
gradient-pairing term of the Fayans functional. At the neutron-rich
side, the difference for the diffraction radii $R$ can amount up to
about 0.1 fm, clearly above the error bars.

\textit{Statistical analysis--}
The information content of $r_4$ is evaluated using standard
statistical correlation analysis as in \cite{Schuetrumpf17,Gor19}. The question we ask is to what
extend $r_4$ is already determined by the other form parameters
and, vice versa, to what extend information on $r_4$ improves our
knowledge of $R$ and $\sigma$. The answer can be quantified in terms 
of statistical correlations. Those between two observables  $A$ and  $B$ are described by  the coefficient of determination CoD($A,B$) deduced from the covariance measure. 
Furthermore, we inspect  multiple correlation coefficients
MCC($A_1,...,A_n;B$), which characterize the correlations between a
group of observables $A_1,...,A_n$ with $B$
\cite{Allison}. The MCC is reduced to the CoD if $n=1$.
 The CoDs and MCCs range from 0 to 1, where 0 implies that the observable $B$ is uncorrelated with the group of observables $A_i$ and the value of 1 means  full correlation.

\begin{figure}
\centerline{\includegraphics[width=0.9\columnwidth]{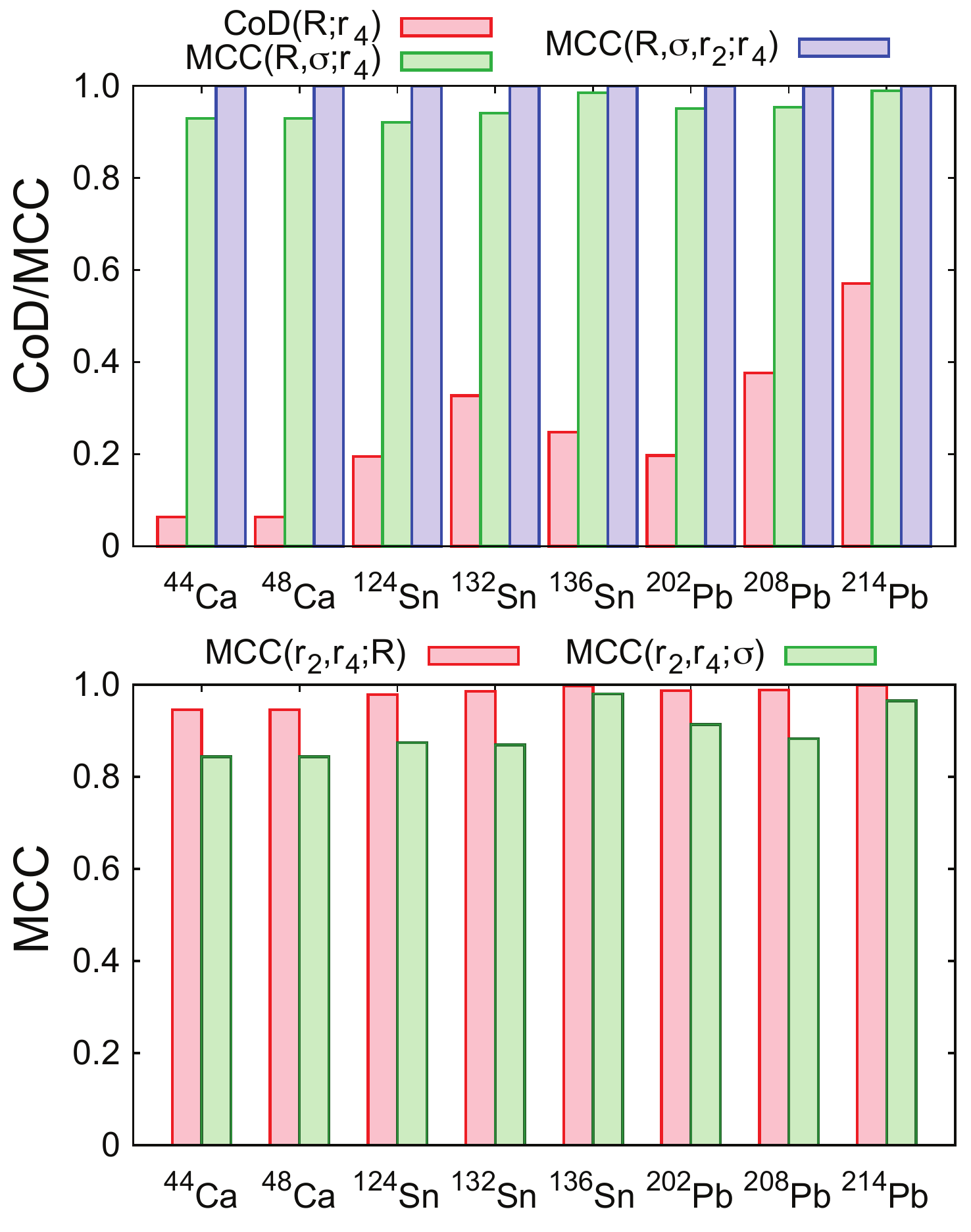}}
\caption{\label{fig:MCC-rad4} 
Top: CoD and MCC coefficients between the standard
form parameters and $r_4$ for a selection of spherical nuclei.
Bottom: information content of $r_2$ and  $r_4$ in terms
of MCCs between $r$ and $r_4$ with the form parameters $R$ or $\sigma$.
}
\end{figure}
The results of our correlation analysis are shown in Fig.~\ref{fig:MCC-rad4}.  The upper panel explores the prediction of $r_4$ for known $R$, $\sigma$, and $r_2$ for a set of spherical nuclei, which have very small halo.
The diffraction radius  alone has little predictive value for $r_4$. This is not
surprising as $R$ carries no information on surface diffuseness, which
 strongly impacts the fourth radial moment.  The combination of
$R$ and $\sigma$  provides  a very good 95\% estimate
of $r_4$. Finally, the group of $R$ and $\sigma$,  and $r_2$ (or $h$)
manages to
determine $r_4$ fully.

The lower panel of Fig.~\ref{fig:MCC-rad4} explores whether a
simultaneous measurement of $r_2$ and $r_4$ can determine $R$ or
$\sigma$. The MCCs show that the diffraction radius is indeed very well
determined by $r_2$ and $r_4$, especially for  heavy nuclei. The surface thickness is also  well predicted, although not as perfectly as $R$, typically at a 90\% level. 

\textit{Helm model analysis--}
Statistical analysis, although well defined and extremely useful, remains
largely a black box. To gain more physics insights, we study  interrelations between the form parameters by virtue of the Helm model.  Given  $r_2$ and $r_4$, we
deduce closed approximate expression for  $\sigma$ and
$R$, again denoted by an upper index (H) to distinguish them from the
exact values, see \cite{SM} for details. For
more compact expressions we introduce the rescaled geometric radii as
$R_n^\mathrm{(g)}= \sqrt[n]{\frac{n+3}{3}r_n^n}$ \cite{Mizutori2000}.

In terms of $R_2^\mathrm{(g)}$ and $R_4^\mathrm{(g)}$  the Helm-model 
values of diffraction radius and surface thickness are \cite{SM}:
\begin{eqnarray}
  R^{\rm (H)}
  &=&\displaystyle
  \sqrt[4]{\frac{7}{2}{R_2^\mathrm{(g)}}_{\mbox{}}^4
           -\frac{5}{2}{R_4^\mathrm{(g)}}_{\mbox{}}^4}, \label{Rh}
\\
  \sigma^{\rm (H)}
  &=& \displaystyle
    \sqrt{\frac{1}{5}{R_2^\mathrm{(g)}}_{\mbox{}}^2
  \left(1
  -
  \sqrt{1
    -\frac{5}{2}\,
     \frac{{R_4^\mathrm{(g)}}_{\mbox{}}^4-{R_2^\mathrm{(g)}}_{\mbox{}}^4}
          {{R_2^\mathrm{(g)}}_{\mbox{}}^4}}
  \right)}.\label{sh}
\end{eqnarray}

Another set of useful relations can be obtained by noticing that $\sigma^2/R^2$ is a small parameter, which is around 0.02-0.03 (see Fig.~\ref{fig:radii}).
By linearizing the above relations with respect
to $\sigma^2/R^2$ one obtains the following approximate relations:
\begin{eqnarray}
 R^{\mathrm{(H),lin}}
  &\approx&
  R_2^\mathrm{(g)}
  +
  \frac{7}{2}\left(R_4^\mathrm{(g)}-R_2^\mathrm{(g)}\right), \label{Rhl}\\
 \sigma^{\mathrm{(H),lin}}
  &\approx&
    \sqrt{\left(R_4^\mathrm{(g)}-R_2^\mathrm{(g)}\right)R_2^\mathrm{(g)}},
\label{shl}\\
 \displaystyle
  \frac{R_4^\mathrm{(g)}-R_2^\mathrm{(g)}}{R_2^\mathrm{(g)}}
  &\approx&
   \displaystyle \left(\frac{\sigma^{\mathrm{(H),lin}}}
   {R^{\mathrm{(H),lin}}}
   \right)^2.
\end{eqnarray}

\begin{figure}
\centerline{\includegraphics[width=0.9\linewidth]{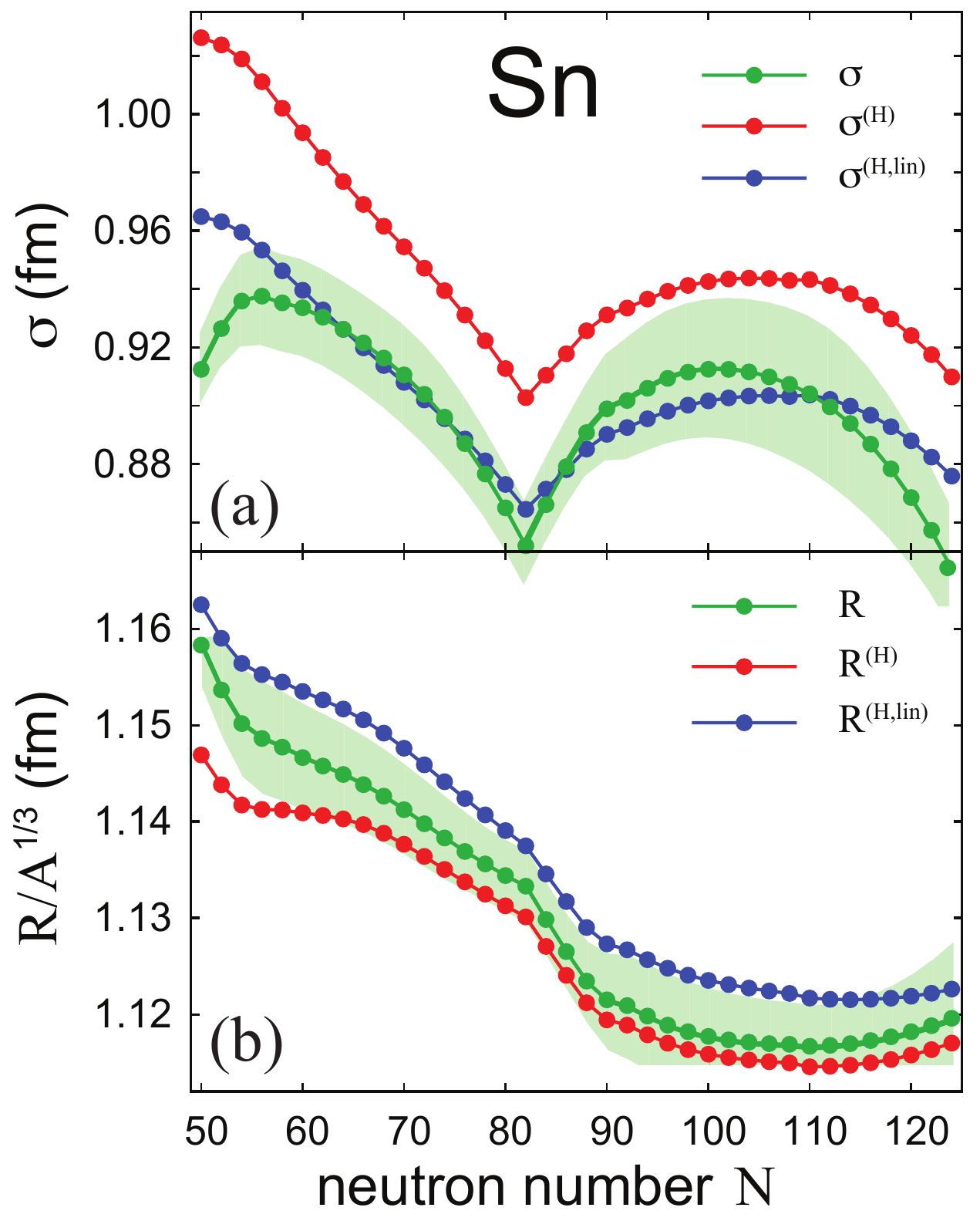}}
\caption{\label{fig:predict-rdifr-Sn-short} Helm-model SV-min predictions
  (\ref{Rh}-\ref{shl}) for $\sigma$ (top) and $R$ (bottom)  for the chain of Sn
  isotopes. The 
   values of
  $\sigma$ and $R$ extracted from the charge density form factor are also shown together with their uncertainties.}
\end{figure}

\begin{figure*}
\centerline{\includegraphics[width=0.9\linewidth]{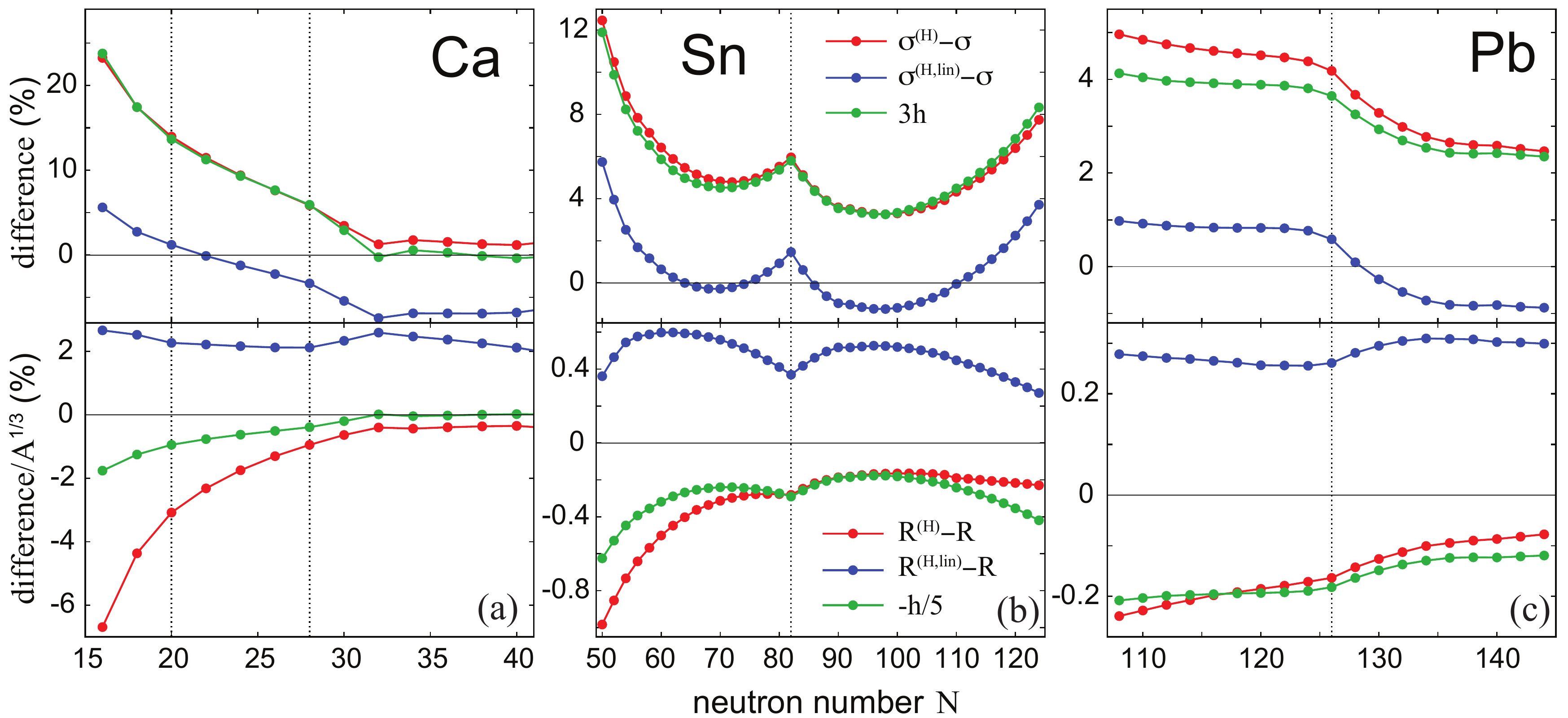}}
\caption{\label{fig:predict-reldiffer-SVmin2} 
 Relative difference of
  Helm-model predictions  (\ref{Rh}-\ref{shl}) from the  form factor values of $R$
  (lower) and $\sigma$ (upper) for the isotopic chains of Ca (a), Sn (b), and Pb (c) computed with SV-min. 
}
\end{figure*}

Figure \ref{fig:predict-rdifr-Sn-short} compares the Helm model values of $R$ and $\sigma$ given by Eqs.~(\ref{Rh}-\ref{shl}) to the exact values directly  obtained form the charge density form factor in the Sn isotopic chain (for Ca and Pb chains, see \cite{SM}). The predictions based on  $r_2$ and $r_4$  are fairly accurate. Indeed for both $R$  and $\sigma$ the deviations of the Helm estimates from the form factor values are close to the   computed uncertainties. Particularly good is the agreement for $R$ as the deviation between  $R^{\mathrm{(H)}}$ and $R$, around 0.02\,fm is smaller than the adopted error of diffraction radii (0.04 fm). 
Interestingly, the linearized $ \sigma^{\mathrm{(H),lin}}$ 
performs exceptionally well except for the most proton-rich isotopes, in which   the appreciable halo feature appears.

Figure \ref{fig:predict-reldiffer-SVmin2} shows the relative differences (in \%)
between form parameters  $R$ and  $\sigma$ and the Helm-model predictions  for three different magic chains: Ca, Sn, and Pb (see \cite{SM} for the absolute differences in fm, and for additional information on $(\sigma/R)^2$). It is seen that the quality of the Helm model predictions improves significantly with increasing system size (see Ref.~\cite{Lept}).  The density
distributions of Ca isotopes are strongly impacted by surface effects and thus harder to
describe by the simple Helm parametrization, while the Pb isotopes are volume-dominated; hence,  they are well
approximated by the Helm model. But the general features observed
before for the Sn chain remain: $R$ is better predicted than $\sigma$,
and the linearized prediction for $\sigma$ performs unexpectedly well
for all isotopic chains (though at different levels of overall
quality).

Altogether, we see that the Helm-model analysis nicely corroborates
the findings from statistical analysis.  Figure~\ref{fig:predict-reldiffer-SVmin2} contains one more piece of
information: it compares the differences with multiples of the halo parameter
(\ref{eq:halo}). It is interesting to see that the  differences
 $R^{\rm (H)}-R$ and 
$ \sigma^{\rm (H)}-\sigma$ are proportional to $h$.
This observation is fairly consistent within the model. Zero halo means that the
exact distribution is described fully by the Helm model. In this case,
the differences would be zero. A mismatch leads to a finite halo and
the same mismatch propagates to the predictions. This suggests a way
to develop model-corrected predictions for $R$ and $\sigma$ from $r_n$
measurements. For a given isotopic chain, the halos follow regular
trends which can be tracked in density-functional calculations. Namely, one can  correct
the Helm-model predictions by 
predicted  halo values to  obtain the form parameters  $R$
and $\sigma$ in exotic nuclei from $r_n$ measurements with well
defined uncertainties that are well below those of the
Helm model.

\textit{Conclusions--}
In this study we assessed the impact of precise experimental determination of $\rf$ on nuclear structure research. 
By means of statistical correlation analysis, we demonstrated that  the diffraction radius
and surface thickness  are  well
determined by $\rt$ and $\rf$, especially for  heavy nuclei. 
This suggests that for those nuclei for which 
precise values of $R$, $\sigma$ and $r_2$ are available,  the values of $r_4$ are fully constrained. Therefore, reliable predictions of these radial moments can be obtained. This will allow realistic estimates of nuclear structure corrections for the interpretation of new physics searches in precision isotope shift measurements.

Experimental determination of $\rf$ would be
extremely valuable for nuclei where the form factor data are not available.  Since electron scattering experiments on unstable nuclei
are highly demanding, if not impossible,  precise measurements of atomic transitions would offer an alternative path to surface properties of unstable nuclei.
Finally, as shown in Fig.~\ref{fig:radii}, information on $\rf$ would be very useful for better constraining current energy density functionals.

\begin{acknowledgements}
This material is based upon work supported by
the U.S. Department of Energy, Office of Science, Office of
Nuclear Physics under Awards No.DE-SC0013365 (Michigan
State University) and DE-SC0018083 (NUCLEI SciDAC-4 collaboration). 
\end{acknowledgements}

\bibliographystyle{apsrev4-1} 
\bibliography{r4}

\begin{thebibliography}{32}%
\makeatletter
\providecommand \@ifxundefined [1]{%
 \@ifx{#1\undefined}
}%
\providecommand \@ifnum [1]{%
 \ifnum #1\expandafter \@firstoftwo
 \else \expandafter \@secondoftwo
 \fi
}%
\providecommand \@ifx [1]{%
 \ifx #1\expandafter \@firstoftwo
 \else \expandafter \@secondoftwo
 \fi
}%
\providecommand \natexlab [1]{#1}%
\providecommand \enquote  [1]{``#1''}%
\providecommand \bibnamefont  [1]{#1}%
\providecommand \bibfnamefont [1]{#1}%
\providecommand \citenamefont [1]{#1}%
\providecommand \href@noop [0]{\@secondoftwo}%
\providecommand \href [0]{\begingroup \@sanitize@url \@href}%
\providecommand \@href[1]{\@@startlink{#1}\@@href}%
\providecommand \@@href[1]{\endgroup#1\@@endlink}%
\providecommand \@sanitize@url [0]{\catcode `\\12\catcode `\$12\catcode
  `\&12\catcode `\#12\catcode `\^12\catcode `\_12\catcode `\%12\relax}%
\providecommand \@@startlink[1]{}%
\providecommand \@@endlink[0]{}%
\providecommand \url  [0]{\begingroup\@sanitize@url \@url }%
\providecommand \@url [1]{\endgroup\@href {#1}{\urlprefix }}%
\providecommand \urlprefix  [0]{URL }%
\providecommand \Eprint [0]{\href }%
\providecommand \doibase [0]{http://dx.doi.org/}%
\providecommand \selectlanguage [0]{\@gobble}%
\providecommand \bibinfo  [0]{\@secondoftwo}%
\providecommand \bibfield  [0]{\@secondoftwo}%
\providecommand \translation [1]{[#1]}%
\providecommand \BibitemOpen [0]{}%
\providecommand \bibitemStop [0]{}%
\providecommand \bibitemNoStop [0]{.\EOS\space}%
\providecommand \EOS [0]{\spacefactor3000\relax}%
\providecommand \BibitemShut  [1]{\csname bibitem#1\endcsname}%
\let\auto@bib@innerbib\@empty
\bibitem [{\citenamefont {Safronova}\ \emph {et~al.}(2018)\citenamefont
  {Safronova}, \citenamefont {Budker}, \citenamefont {DeMille}, \citenamefont
  {Kimball}, \citenamefont {Derevianko},\ and\ \citenamefont {Clark}}]{Saf18}%
  \BibitemOpen
  \bibfield  {author} {\bibinfo {author} {\bibfnamefont {M.~S.}\ \bibnamefont
  {Safronova}}, \bibinfo {author} {\bibfnamefont {D.}~\bibnamefont {Budker}},
  \bibinfo {author} {\bibfnamefont {D.}~\bibnamefont {DeMille}}, \bibinfo
  {author} {\bibfnamefont {D.~F.~J.}\ \bibnamefont {Kimball}}, \bibinfo
  {author} {\bibfnamefont {A.}~\bibnamefont {Derevianko}}, \ and\ \bibinfo
  {author} {\bibfnamefont {C.~W.}\ \bibnamefont {Clark}},\ }\href {\doibase
  10.1103/RevModPhys.90.025008} {\bibfield  {journal} {\bibinfo  {journal}
  {Rev. Mod. Phys.}\ }\textbf {\bibinfo {volume} {90}},\ \bibinfo {pages}
  {025008} (\bibinfo {year} {2018})}\BibitemShut {NoStop}%
\bibitem [{\citenamefont {Berengut}\ \emph {et~al.}(2018)\citenamefont
  {Berengut} \emph {et~al.}}]{Ber18}%
  \BibitemOpen
  \bibfield  {author} {\bibinfo {author} {\bibfnamefont {J.~C.}\ \bibnamefont
  {Berengut}} \emph {et~al.},\ }\href {\doibase 10.1103/PhysRevLett.120.091801}
  {\bibfield  {journal} {\bibinfo  {journal} {Phys. Rev. Lett.}\ }\textbf
  {\bibinfo {volume} {120}},\ \bibinfo {pages} {91801} (\bibinfo {year}
  {2018})}\BibitemShut {NoStop}%
\bibitem [{\citenamefont {Stadnik}(2018)}]{Sta18}%
  \BibitemOpen
  \bibfield  {author} {\bibinfo {author} {\bibfnamefont {Y.~V.}\ \bibnamefont
  {Stadnik}},\ }\href {\doibase 10.1103/PhysRevLett.120.223202} {\bibfield
  {journal} {\bibinfo  {journal} {Phys. Rev. Lett.}\ }\textbf {\bibinfo
  {volume} {120}},\ \bibinfo {pages} {223202} (\bibinfo {year}
  {2018})}\BibitemShut {NoStop}%
\bibitem [{\citenamefont {Delaunay}\ \emph
  {et~al.}(2017{\natexlab{a}})\citenamefont {Delaunay}, \citenamefont
  {Frugiuele}, \citenamefont {Fuchs},\ and\ \citenamefont {Soreq}}]{Del17}%
  \BibitemOpen
  \bibfield  {author} {\bibinfo {author} {\bibfnamefont {C.}~\bibnamefont
  {Delaunay}}, \bibinfo {author} {\bibfnamefont {C.}~\bibnamefont {Frugiuele}},
  \bibinfo {author} {\bibfnamefont {E.}~\bibnamefont {Fuchs}}, \ and\ \bibinfo
  {author} {\bibfnamefont {Y.}~\bibnamefont {Soreq}},\ }\href {\doibase
  10.1103/PhysRevD.96.115002} {\bibfield  {journal} {\bibinfo  {journal} {Phys.
  Rev. D}\ }\textbf {\bibinfo {volume} {96}},\ \bibinfo {pages} {115002}
  (\bibinfo {year} {2017}{\natexlab{a}})}\BibitemShut {NoStop}%
\bibitem [{\citenamefont {{Garcia Ruiz}}\ \emph {et~al.}(2016)\citenamefont
  {{Garcia Ruiz}} \emph {et~al.}}]{Gar16}%
  \BibitemOpen
  \bibfield  {author} {\bibinfo {author} {\bibfnamefont {R.~F.}\ \bibnamefont
  {{Garcia Ruiz}}} \emph {et~al.},\ }\href {\doibase 10.1038/nphys3645}
  {\bibfield  {journal} {\bibinfo  {journal} {Nature Phys.}\ }\textbf {\bibinfo
  {volume} {12}},\ \bibinfo {pages} {594} (\bibinfo {year} {2016})}\BibitemShut
  {NoStop}%
\bibitem [{\citenamefont {Campbell}\ \emph {et~al.}(2016)\citenamefont
  {Campbell}, \citenamefont {Moore},\ and\ \citenamefont {Pearson}}]{Cam16}%
  \BibitemOpen
  \bibfield  {author} {\bibinfo {author} {\bibfnamefont {P.}~\bibnamefont
  {Campbell}}, \bibinfo {author} {\bibfnamefont {I.}~\bibnamefont {Moore}}, \
  and\ \bibinfo {author} {\bibfnamefont {M.}~\bibnamefont {Pearson}},\ }\href
  {\doibase 10.1016/j.ppnp.2015.09.003} {\bibfield  {journal} {\bibinfo
  {journal} {Progress in Particle and Nuclear Physics}\ }\textbf {\bibinfo
  {volume} {86}},\ \bibinfo {pages} {127} (\bibinfo {year} {2016})}\BibitemShut
  {NoStop}%
\bibitem [{\citenamefont {Hammen}\ \emph {et~al.}(2018)\citenamefont {Hammen},
  \citenamefont {N{\"{o}}rtersh{\"{a}}user}, \citenamefont {Balabanski},
  \citenamefont {Bissell}, \citenamefont {Blaum}, \citenamefont {Cheal},
  \citenamefont {Flanagan}, \citenamefont {Fr{\"{o}}mmgen}, \citenamefont
  {Georgiev}, \citenamefont {Geppert}, \citenamefont {Kowalska}, \citenamefont
  {Kreim}, \citenamefont {Krieger}, \citenamefont {Nazarewicz}, \citenamefont
  {Neugart}, \citenamefont {Neyens}, \citenamefont {Papuga}, \citenamefont
  {Reinhard}, \citenamefont {Rajabali}, \citenamefont {Schmidt},\ and\
  \citenamefont {Yordanov}}]{Ham18}%
  \BibitemOpen
  \bibfield  {author} {\bibinfo {author} {\bibfnamefont {M.}~\bibnamefont
  {Hammen}}, \bibinfo {author} {\bibfnamefont {W.}~\bibnamefont
  {N{\"{o}}rtersh{\"{a}}user}}, \bibinfo {author} {\bibfnamefont {D.~L.}\
  \bibnamefont {Balabanski}}, \bibinfo {author} {\bibfnamefont {M.~L.}\
  \bibnamefont {Bissell}}, \bibinfo {author} {\bibfnamefont {K.}~\bibnamefont
  {Blaum}}, \bibinfo {author} {\bibfnamefont {B.}~\bibnamefont {Cheal}},
  \bibinfo {author} {\bibfnamefont {K.~T.}\ \bibnamefont {Flanagan}}, \bibinfo
  {author} {\bibfnamefont {N.}~\bibnamefont {Fr{\"{o}}mmgen}}, \bibinfo
  {author} {\bibfnamefont {G.}~\bibnamefont {Georgiev}}, \bibinfo {author}
  {\bibfnamefont {C.}~\bibnamefont {Geppert}}, \bibinfo {author} {\bibfnamefont
  {M.}~\bibnamefont {Kowalska}}, \bibinfo {author} {\bibfnamefont
  {K.}~\bibnamefont {Kreim}}, \bibinfo {author} {\bibfnamefont
  {A.}~\bibnamefont {Krieger}}, \bibinfo {author} {\bibfnamefont
  {W.}~\bibnamefont {Nazarewicz}}, \bibinfo {author} {\bibfnamefont
  {R.}~\bibnamefont {Neugart}}, \bibinfo {author} {\bibfnamefont
  {G.}~\bibnamefont {Neyens}}, \bibinfo {author} {\bibfnamefont
  {J.}~\bibnamefont {Papuga}}, \bibinfo {author} {\bibfnamefont {P.-G.}\
  \bibnamefont {Reinhard}}, \bibinfo {author} {\bibfnamefont {M.~M.}\
  \bibnamefont {Rajabali}}, \bibinfo {author} {\bibfnamefont {S.}~\bibnamefont
  {Schmidt}}, \ and\ \bibinfo {author} {\bibfnamefont {D.~T.}\ \bibnamefont
  {Yordanov}},\ }\href {\doibase 10.1103/PhysRevLett.121.102501} {\bibfield
  {journal} {\bibinfo  {journal} {Phys. Rev. Lett.}\ }\textbf {\bibinfo
  {volume} {121}},\ \bibinfo {pages} {102501} (\bibinfo {year}
  {2018})}\BibitemShut {NoStop}%
\bibitem [{\citenamefont {Marsh}\ \emph {et~al.}(2018)\citenamefont {Marsh}
  \emph {et~al.}}]{Marsh18}%
  \BibitemOpen
  \bibfield  {author} {\bibinfo {author} {\bibfnamefont {B.~A.}\ \bibnamefont
  {Marsh}} \emph {et~al.},\ }\href {\doibase 10.1038/s41567-018-0292-8}
  {\bibfield  {journal} {\bibinfo  {journal} {Nature Phys.}\ }\textbf {\bibinfo
  {volume} {14}},\ \bibinfo {pages} {1163} (\bibinfo {year}
  {2018})}\BibitemShut {NoStop}%
\bibitem [{\citenamefont {Gorges}\ \emph {et~al.}(2019)\citenamefont {Gorges},
  \citenamefont {Rodr\'{\i}guez}, \citenamefont {Balabanski}, \citenamefont
  {Bissell}, \citenamefont {Blaum}, \citenamefont {Cheal}, \citenamefont
  {Garcia~Ruiz}, \citenamefont {Georgiev}, \citenamefont {Gins}, \citenamefont
  {Heylen}, \citenamefont {Kanellakopoulos}, \citenamefont {Kaufmann},
  \citenamefont {Kowalska}, \citenamefont {Lagaki}, \citenamefont {Lechner},
  \citenamefont {Maa\ss{}}, \citenamefont {Malbrunot-Ettenauer}, \citenamefont
  {Nazarewicz}, \citenamefont {Neugart}, \citenamefont {Neyens}, \citenamefont
  {N\"ortersh\"auser}, \citenamefont {Reinhard}, \citenamefont {Sailer},
  \citenamefont {S\'anchez}, \citenamefont {Schmidt}, \citenamefont {Wehner},
  \citenamefont {Wraith}, \citenamefont {Xie}, \citenamefont {Xu},
  \citenamefont {Yang},\ and\ \citenamefont {Yordanov}}]{Gor19}%
  \BibitemOpen
  \bibfield  {author} {\bibinfo {author} {\bibfnamefont {C.}~\bibnamefont
  {Gorges}}, \bibinfo {author} {\bibfnamefont {L.~V.}\ \bibnamefont
  {Rodr\'{\i}guez}}, \bibinfo {author} {\bibfnamefont {D.~L.}\ \bibnamefont
  {Balabanski}}, \bibinfo {author} {\bibfnamefont {M.~L.}\ \bibnamefont
  {Bissell}}, \bibinfo {author} {\bibfnamefont {K.}~\bibnamefont {Blaum}},
  \bibinfo {author} {\bibfnamefont {B.}~\bibnamefont {Cheal}}, \bibinfo
  {author} {\bibfnamefont {R.~F.}\ \bibnamefont {Garcia~Ruiz}}, \bibinfo
  {author} {\bibfnamefont {G.}~\bibnamefont {Georgiev}}, \bibinfo {author}
  {\bibfnamefont {W.}~\bibnamefont {Gins}}, \bibinfo {author} {\bibfnamefont
  {H.}~\bibnamefont {Heylen}}, \bibinfo {author} {\bibfnamefont
  {A.}~\bibnamefont {Kanellakopoulos}}, \bibinfo {author} {\bibfnamefont
  {S.}~\bibnamefont {Kaufmann}}, \bibinfo {author} {\bibfnamefont
  {M.}~\bibnamefont {Kowalska}}, \bibinfo {author} {\bibfnamefont
  {V.}~\bibnamefont {Lagaki}}, \bibinfo {author} {\bibfnamefont
  {S.}~\bibnamefont {Lechner}}, \bibinfo {author} {\bibfnamefont
  {B.}~\bibnamefont {Maa\ss{}}}, \bibinfo {author} {\bibfnamefont
  {S.}~\bibnamefont {Malbrunot-Ettenauer}}, \bibinfo {author} {\bibfnamefont
  {W.}~\bibnamefont {Nazarewicz}}, \bibinfo {author} {\bibfnamefont
  {R.}~\bibnamefont {Neugart}}, \bibinfo {author} {\bibfnamefont
  {G.}~\bibnamefont {Neyens}}, \bibinfo {author} {\bibfnamefont
  {W.}~\bibnamefont {N\"ortersh\"auser}}, \bibinfo {author} {\bibfnamefont
  {P.-G.}\ \bibnamefont {Reinhard}}, \bibinfo {author} {\bibfnamefont
  {S.}~\bibnamefont {Sailer}}, \bibinfo {author} {\bibfnamefont
  {R.}~\bibnamefont {S\'anchez}}, \bibinfo {author} {\bibfnamefont
  {S.}~\bibnamefont {Schmidt}}, \bibinfo {author} {\bibfnamefont
  {L.}~\bibnamefont {Wehner}}, \bibinfo {author} {\bibfnamefont
  {C.}~\bibnamefont {Wraith}}, \bibinfo {author} {\bibfnamefont
  {L.}~\bibnamefont {Xie}}, \bibinfo {author} {\bibfnamefont {Z.~Y.}\
  \bibnamefont {Xu}}, \bibinfo {author} {\bibfnamefont {X.~F.}\ \bibnamefont
  {Yang}}, \ and\ \bibinfo {author} {\bibfnamefont {D.~T.}\ \bibnamefont
  {Yordanov}},\ }\href {\doibase 10.1103/PhysRevLett.122.192502} {\bibfield
  {journal} {\bibinfo  {journal} {Phys. Rev. Lett.}\ }\textbf {\bibinfo
  {volume} {122}},\ \bibinfo {pages} {192502} (\bibinfo {year}
  {2019})}\BibitemShut {NoStop}%
\bibitem [{\citenamefont {{Miller}}\ \emph {et~al.}(2019)\citenamefont
  {{Miller}} \emph {et~al.}}]{Mil19}%
  \BibitemOpen
  \bibfield  {author} {\bibinfo {author} {\bibfnamefont {A.~J.}\ \bibnamefont
  {{Miller}}} \emph {et~al.},\ }\href {\doibase 10.1038/s41567-019-0416-9}
  {\bibfield  {journal} {\bibinfo  {journal} {Nature Phys.}\ }\textbf {\bibinfo
  {volume} {15}},\ \bibinfo {pages} {432} (\bibinfo {year} {2019})}\BibitemShut
  {NoStop}%
\bibitem [{\citenamefont {Delaunay}\ \emph
  {et~al.}(2017{\natexlab{b}})\citenamefont {Delaunay}, \citenamefont {Ozeri},
  \citenamefont {Perez},\ and\ \citenamefont {Soreq}}]{Del17b}%
  \BibitemOpen
  \bibfield  {author} {\bibinfo {author} {\bibfnamefont {C.}~\bibnamefont
  {Delaunay}}, \bibinfo {author} {\bibfnamefont {R.}~\bibnamefont {Ozeri}},
  \bibinfo {author} {\bibfnamefont {G.}~\bibnamefont {Perez}}, \ and\ \bibinfo
  {author} {\bibfnamefont {Y.}~\bibnamefont {Soreq}},\ }\href {\doibase
  10.1103/PhysRevD.96.093001} {\bibfield  {journal} {\bibinfo  {journal} {Phys.
  Rev. D}\ }\textbf {\bibinfo {volume} {96}},\ \bibinfo {pages} {93001}
  (\bibinfo {year} {2017}{\natexlab{b}})}\BibitemShut {NoStop}%
\bibitem [{\citenamefont {Frugiuele}\ \emph {et~al.}(2017)\citenamefont
  {Frugiuele}, \citenamefont {Fuchs}, \citenamefont {Perez},\ and\
  \citenamefont {Schlaffer}}]{Fru17}%
  \BibitemOpen
  \bibfield  {author} {\bibinfo {author} {\bibfnamefont {C.}~\bibnamefont
  {Frugiuele}}, \bibinfo {author} {\bibfnamefont {E.}~\bibnamefont {Fuchs}},
  \bibinfo {author} {\bibfnamefont {G.}~\bibnamefont {Perez}}, \ and\ \bibinfo
  {author} {\bibfnamefont {M.}~\bibnamefont {Schlaffer}},\ }\href {\doibase
  10.1103/PhysRevD.96.015011} {\bibfield  {journal} {\bibinfo  {journal} {Phys.
  Rev. D}\ }\textbf {\bibinfo {volume} {96}},\ \bibinfo {pages} {015011}
  (\bibinfo {year} {2017})}\BibitemShut {NoStop}%
\bibitem [{\citenamefont {Flambaum}\ \emph {et~al.}(2018)\citenamefont
  {Flambaum}, \citenamefont {Geddes},\ and\ \citenamefont {Viatkina}}]{Fla18}%
  \BibitemOpen
  \bibfield  {author} {\bibinfo {author} {\bibfnamefont {V.~V.}\ \bibnamefont
  {Flambaum}}, \bibinfo {author} {\bibfnamefont {A.~J.}\ \bibnamefont
  {Geddes}}, \ and\ \bibinfo {author} {\bibfnamefont {A.~V.}\ \bibnamefont
  {Viatkina}},\ }\href {\doibase 10.1103/PhysRevA.97.032510} {\bibfield
  {journal} {\bibinfo  {journal} {Phys. Rev. A}\ }\textbf {\bibinfo {volume}
  {97}},\ \bibinfo {pages} {032510} (\bibinfo {year} {2018})}\BibitemShut
  {NoStop}%
\bibitem [{\citenamefont {Gebert}\ \emph {et~al.}(2015)\citenamefont {Gebert},
  \citenamefont {Wan}, \citenamefont {Wolf}, \citenamefont {Angstmann},
  \citenamefont {Berengut},\ and\ \citenamefont {Schmidt}}]{Geb15}%
  \BibitemOpen
  \bibfield  {author} {\bibinfo {author} {\bibfnamefont {F.}~\bibnamefont
  {Gebert}}, \bibinfo {author} {\bibfnamefont {Y.}~\bibnamefont {Wan}},
  \bibinfo {author} {\bibfnamefont {F.}~\bibnamefont {Wolf}}, \bibinfo {author}
  {\bibfnamefont {C.~N.}\ \bibnamefont {Angstmann}}, \bibinfo {author}
  {\bibfnamefont {J.~C.}\ \bibnamefont {Berengut}}, \ and\ \bibinfo {author}
  {\bibfnamefont {P.~O.}\ \bibnamefont {Schmidt}},\ }\href {\doibase
  10.1103/PhysRevLett.115.053003} {\bibfield  {journal} {\bibinfo  {journal}
  {Phys. Rev. Lett.}\ }\textbf {\bibinfo {volume} {115}},\ \bibinfo {pages}
  {053003} (\bibinfo {year} {2015})}\BibitemShut {NoStop}%
\bibitem [{\citenamefont {Braverman}\ \emph {et~al.}(2019)\citenamefont
  {Braverman}, \citenamefont {Kawasaki}, \citenamefont {Pedrozo-Pe\~nafiel},
  \citenamefont {Colombo}, \citenamefont {Shu}, \citenamefont {Li},
  \citenamefont {Mendez}, \citenamefont {Yamoah}, \citenamefont {Salvi},
  \citenamefont {Akamatsu}, \citenamefont {Xiao},\ and\ \citenamefont
  {Vuleti\ifmmode~\acute{c}\else \'{c}\fi{}}}]{Bra19}%
  \BibitemOpen
  \bibfield  {author} {\bibinfo {author} {\bibfnamefont {B.}~\bibnamefont
  {Braverman}}, \bibinfo {author} {\bibfnamefont {A.}~\bibnamefont {Kawasaki}},
  \bibinfo {author} {\bibfnamefont {E.}~\bibnamefont {Pedrozo-Pe\~nafiel}},
  \bibinfo {author} {\bibfnamefont {S.}~\bibnamefont {Colombo}}, \bibinfo
  {author} {\bibfnamefont {C.}~\bibnamefont {Shu}}, \bibinfo {author}
  {\bibfnamefont {Z.}~\bibnamefont {Li}}, \bibinfo {author} {\bibfnamefont
  {E.}~\bibnamefont {Mendez}}, \bibinfo {author} {\bibfnamefont
  {M.}~\bibnamefont {Yamoah}}, \bibinfo {author} {\bibfnamefont
  {L.}~\bibnamefont {Salvi}}, \bibinfo {author} {\bibfnamefont
  {D.}~\bibnamefont {Akamatsu}}, \bibinfo {author} {\bibfnamefont
  {Y.}~\bibnamefont {Xiao}}, \ and\ \bibinfo {author} {\bibfnamefont
  {V.}~\bibnamefont {Vuleti\ifmmode~\acute{c}\else \'{c}\fi{}}},\ }\href
  {\doibase 10.1103/PhysRevLett.122.223203} {\bibfield  {journal} {\bibinfo
  {journal} {Phys. Rev. Lett.}\ }\textbf {\bibinfo {volume} {122}},\ \bibinfo
  {pages} {223203} (\bibinfo {year} {2019})}\BibitemShut {NoStop}%
\bibitem [{\citenamefont {Manovitz}\ \emph {et~al.}(2019)\citenamefont
  {Manovitz} \emph {et~al.}}]{Man19}%
  \BibitemOpen
  \bibfield  {author} {\bibinfo {author} {\bibfnamefont {T.}~\bibnamefont
  {Manovitz}} \emph {et~al.},\ }\href@noop {} {\bibfield  {journal} {\bibinfo
  {journal} {arxiv.org/abs/1906.05770v1}\ } (\bibinfo {year}
  {2019})}\BibitemShut {NoStop}%
\bibitem [{\citenamefont {Papoulia}\ \emph {et~al.}(2016)\citenamefont
  {Papoulia}, \citenamefont {Carlsson},\ and\ \citenamefont {Ekman}}]{Pap16}%
  \BibitemOpen
  \bibfield  {author} {\bibinfo {author} {\bibfnamefont {A.}~\bibnamefont
  {Papoulia}}, \bibinfo {author} {\bibfnamefont {B.~G.}\ \bibnamefont
  {Carlsson}}, \ and\ \bibinfo {author} {\bibfnamefont {J.}~\bibnamefont
  {Ekman}},\ }\href {\doibase 10.1103/PhysRevA.94.042502} {\bibfield  {journal}
  {\bibinfo  {journal} {Phys. Rev. A}\ }\textbf {\bibinfo {volume} {94}},\
  \bibinfo {pages} {042502} (\bibinfo {year} {2016})}\BibitemShut {NoStop}%
\bibitem [{\citenamefont {Ekman}\ \emph {et~al.}(2019)\citenamefont {Ekman},
  \citenamefont {J{\"o}nsson}, \citenamefont {Godefroid}, \citenamefont
  {Naz{\'e}}, \citenamefont {Gaigalas},\ and\ \citenamefont
  {Biero{\'n}}}]{Ekman2019}%
  \BibitemOpen
  \bibfield  {author} {\bibinfo {author} {\bibfnamefont {J.}~\bibnamefont
  {Ekman}}, \bibinfo {author} {\bibfnamefont {P.}~\bibnamefont {J{\"o}nsson}},
  \bibinfo {author} {\bibfnamefont {M.}~\bibnamefont {Godefroid}}, \bibinfo
  {author} {\bibfnamefont {C.}~\bibnamefont {Naz{\'e}}}, \bibinfo {author}
  {\bibfnamefont {G.}~\bibnamefont {Gaigalas}}, \ and\ \bibinfo {author}
  {\bibfnamefont {J.}~\bibnamefont {Biero{\'n}}},\ }\href {\doibase
  https://doi.org/10.1016/j.cpc.2018.08.017} {\bibfield  {journal} {\bibinfo
  {journal} {Comput. Phys. Commun.}\ }\textbf {\bibinfo {volume} {235}},\
  \bibinfo {pages} {433 } (\bibinfo {year} {2019})}\BibitemShut {NoStop}%
\bibitem [{\citenamefont {King}(2013)}]{king}%
  \BibitemOpen
  \bibfield  {author} {\bibinfo {author} {\bibfnamefont {W.}~\bibnamefont
  {King}},\ }\href {https://books.google.com/books?id=eEgGCAAAQBAJ{\&}pgis=1}
  {\emph {\bibinfo {title} {{Isotope Shifts in Atomic Spectra}}}},\
  Vol.~\bibinfo {volume} {11}\ (\bibinfo  {publisher} {Springer Science {\&}
  Business Media},\ \bibinfo {year} {2013})\ p.\ \bibinfo {pages}
  {210}\BibitemShut {NoStop}%
\bibitem [{\citenamefont {Helm}(1956)}]{Hel56aE}%
  \BibitemOpen
  \bibfield  {author} {\bibinfo {author} {\bibfnamefont {R.~H.}\ \bibnamefont
  {Helm}},\ }\href {\doibase 10.1103/PhysRev.104.1466} {\bibfield  {journal}
  {\bibinfo  {journal} {Phys. Rev.}\ }\textbf {\bibinfo {volume} {104}},\
  \bibinfo {pages} {1466} (\bibinfo {year} {1956})}\BibitemShut {NoStop}%
\bibitem [{\citenamefont {Friedrich}\ and\ \citenamefont
  {V{\"o}gler}(1982)}]{Fri82a}%
  \BibitemOpen
  \bibfield  {author} {\bibinfo {author} {\bibfnamefont {J.}~\bibnamefont
  {Friedrich}}\ and\ \bibinfo {author} {\bibfnamefont {N.}~\bibnamefont
  {V{\"o}gler}},\ }\href {\doibase 10.1016/0375-9474(82)90147-6} {\bibfield
  {journal} {\bibinfo  {journal} {Nucl. Phys. A}\ }\textbf {\bibinfo {volume}
  {373}},\ \bibinfo {pages} {192} (\bibinfo {year} {1982})}\BibitemShut
  {NoStop}%
\bibitem [{\citenamefont {Andrae}(2000)}]{Andrae2000}%
  \BibitemOpen
  \bibfield  {author} {\bibinfo {author} {\bibfnamefont {D.}~\bibnamefont
  {Andrae}},\ }\href {\doibase 10.1016/S0370-1573(00)00007-7} {\bibfield
  {journal} {\bibinfo  {journal} {Phys. Rep.}\ }\textbf {\bibinfo {volume}
  {336}},\ \bibinfo {pages} {413 } (\bibinfo {year} {2000})}\BibitemShut
  {NoStop}%
\bibitem [{SM()}]{SM}%
  \BibitemOpen
  \href@noop {} {}\bibinfo {note} {See Supplemental Material at
  \url{http://link.aps.org/supplemental/XXX}}\BibitemShut {NoStop}%
\bibitem [{\citenamefont {Mizutori}\ \emph {et~al.}(2000)\citenamefont
  {Mizutori}, \citenamefont {Dobaczewski}, \citenamefont {Lalazissis},
  \citenamefont {Nazarewicz},\ and\ \citenamefont {Reinhard}}]{Mizutori2000}%
  \BibitemOpen
  \bibfield  {author} {\bibinfo {author} {\bibfnamefont {S.}~\bibnamefont
  {Mizutori}}, \bibinfo {author} {\bibfnamefont {J.}~\bibnamefont
  {Dobaczewski}}, \bibinfo {author} {\bibfnamefont {G.}~\bibnamefont
  {Lalazissis}}, \bibinfo {author} {\bibfnamefont {W.}~\bibnamefont
  {Nazarewicz}}, \ and\ \bibinfo {author} {\bibfnamefont {P.-G.}\ \bibnamefont
  {Reinhard}},\ }\href {\doibase 10.1103/PhysRevC.61.044326} {\bibfield
  {journal} {\bibinfo  {journal} {Phys. Rev. C}\ }\textbf {\bibinfo {volume}
  {61}},\ \bibinfo {pages} {044326} (\bibinfo {year} {2000})}\BibitemShut
  {NoStop}%
\bibitem [{\citenamefont {Sarriguren}\ \emph {et~al.}(2007)\citenamefont
  {Sarriguren}, \citenamefont {Gaidarov}, \citenamefont {Guerra},\ and\
  \citenamefont {Antonov}}]{Sarriguren07}%
  \BibitemOpen
  \bibfield  {author} {\bibinfo {author} {\bibfnamefont {P.}~\bibnamefont
  {Sarriguren}}, \bibinfo {author} {\bibfnamefont {M.~K.}\ \bibnamefont
  {Gaidarov}}, \bibinfo {author} {\bibfnamefont {E.~M.~d.}\ \bibnamefont
  {Guerra}}, \ and\ \bibinfo {author} {\bibfnamefont {A.~N.}\ \bibnamefont
  {Antonov}},\ }\href {\doibase 10.1103/PhysRevC.76.044322} {\bibfield
  {journal} {\bibinfo  {journal} {Phys. Rev. C}\ }\textbf {\bibinfo {volume}
  {76}},\ \bibinfo {pages} {044322} (\bibinfo {year} {2007})}\BibitemShut
  {NoStop}%
\bibitem [{\citenamefont {Schunck}\ and\ \citenamefont
  {Egido}(2008)}]{Schunck08}%
  \BibitemOpen
  \bibfield  {author} {\bibinfo {author} {\bibfnamefont {N.}~\bibnamefont
  {Schunck}}\ and\ \bibinfo {author} {\bibfnamefont {J.~L.}\ \bibnamefont
  {Egido}},\ }\href {\doibase 10.1103/PhysRevC.78.064305} {\bibfield  {journal}
  {\bibinfo  {journal} {Phys. Rev. C}\ }\textbf {\bibinfo {volume} {78}},\
  \bibinfo {pages} {064305} (\bibinfo {year} {2008})}\BibitemShut {NoStop}%
\bibitem [{\citenamefont {Kl\"upfel}\ \emph {et~al.}(2009)\citenamefont
  {Kl\"upfel}, \citenamefont {Reinhard}, \citenamefont {B\"urvenich},\ and\
  \citenamefont {Maruhn}}]{Klu09a}%
  \BibitemOpen
  \bibfield  {author} {\bibinfo {author} {\bibfnamefont {P.}~\bibnamefont
  {Kl\"upfel}}, \bibinfo {author} {\bibfnamefont {P.-G.}\ \bibnamefont
  {Reinhard}}, \bibinfo {author} {\bibfnamefont {T.~J.}\ \bibnamefont
  {B\"urvenich}}, \ and\ \bibinfo {author} {\bibfnamefont {J.~A.}\ \bibnamefont
  {Maruhn}},\ }\href {\doibase 10.1103/PhysRevC.79.034310} {\bibfield
  {journal} {\bibinfo  {journal} {Phys. Rev. C}\ }\textbf {\bibinfo {volume}
  {79}},\ \bibinfo {pages} {034310} (\bibinfo {year} {2009})}\BibitemShut
  {NoStop}%
\bibitem [{\citenamefont {Reinhard}\ and\ \citenamefont
  {Nazarewicz}(2017)}]{Rei17a}%
  \BibitemOpen
  \bibfield  {author} {\bibinfo {author} {\bibfnamefont {P.-G.}\ \bibnamefont
  {Reinhard}}\ and\ \bibinfo {author} {\bibfnamefont {W.}~\bibnamefont
  {Nazarewicz}},\ }\href {\doibase 10.1103/PhysRevC.95.064328} {\bibfield
  {journal} {\bibinfo  {journal} {Phys. Rev. C}\ }\textbf {\bibinfo {volume}
  {95}},\ \bibinfo {pages} {064328} (\bibinfo {year} {2017})}\BibitemShut
  {NoStop}%
\bibitem [{\citenamefont {Dobaczewski}\ \emph {et~al.}(2014)\citenamefont
  {Dobaczewski}, \citenamefont {Nazarewicz},\ and\ \citenamefont
  {Reinhard}}]{Dob14a}%
  \BibitemOpen
  \bibfield  {author} {\bibinfo {author} {\bibfnamefont {J.}~\bibnamefont
  {Dobaczewski}}, \bibinfo {author} {\bibfnamefont {W.}~\bibnamefont
  {Nazarewicz}}, \ and\ \bibinfo {author} {\bibfnamefont {P.-G.}\ \bibnamefont
  {Reinhard}},\ }\href {\doibase 10.1088/0954-3899/41/7/074001} {\bibfield
  {journal} {\bibinfo  {journal} {J. Phys. G}\ }\textbf {\bibinfo {volume}
  {41}},\ \bibinfo {pages} {074001} (\bibinfo {year} {2014})}\BibitemShut
  {NoStop}%
\bibitem [{\citenamefont {Schuetrumpf}\ \emph {et~al.}(2017)\citenamefont
  {Schuetrumpf}, \citenamefont {Nazarewicz},\ and\ \citenamefont
  {Reinhard}}]{Schuetrumpf17}%
  \BibitemOpen
  \bibfield  {author} {\bibinfo {author} {\bibfnamefont {B.}~\bibnamefont
  {Schuetrumpf}}, \bibinfo {author} {\bibfnamefont {W.}~\bibnamefont
  {Nazarewicz}}, \ and\ \bibinfo {author} {\bibfnamefont {P.-G.}\ \bibnamefont
  {Reinhard}},\ }\href {\doibase 10.1103/PhysRevC.96.024306} {\bibfield
  {journal} {\bibinfo  {journal} {Phys. Rev. C}\ }\textbf {\bibinfo {volume}
  {96}},\ \bibinfo {pages} {024306} (\bibinfo {year} {2017})}\BibitemShut
  {NoStop}%
\bibitem [{\citenamefont {Allison}(1998)}]{Allison}%
  \BibitemOpen
  \bibfield  {author} {\bibinfo {author} {\bibfnamefont {P.~D.}\ \bibnamefont
  {Allison}},\ }\href@noop {} {\emph {\bibinfo {title} {Multiple Regression: A
  Primer}}}\ (\bibinfo  {publisher} {Sage Publications, Thousand Oaks, CA,},\
  \bibinfo {year} {1998})\BibitemShut {NoStop}%
\bibitem [{\citenamefont {Reinhard}\ \emph {et~al.}(2006)\citenamefont
  {Reinhard}, \citenamefont {Bender}, \citenamefont {Nazarewicz},\ and\
  \citenamefont {Vertse}}]{Lept}%
  \BibitemOpen
  \bibfield  {author} {\bibinfo {author} {\bibfnamefont {P.-G.}\ \bibnamefont
  {Reinhard}}, \bibinfo {author} {\bibfnamefont {M.}~\bibnamefont {Bender}},
  \bibinfo {author} {\bibfnamefont {W.}~\bibnamefont {Nazarewicz}}, \ and\
  \bibinfo {author} {\bibfnamefont {T.}~\bibnamefont {Vertse}},\ }\href
  {\doibase 10.1103/PhysRevC.73.014309} {\bibfield  {journal} {\bibinfo
  {journal} {Phys. Rev. C}\ }\textbf {\bibinfo {volume} {73}},\ \bibinfo
  {pages} {014309} (\bibinfo {year} {2006})}\BibitemShut {NoStop}%
\end{thebibliography}%

\end{document}